\documentclass[journal]{IEEEtran}

\usepackage{cite}
\usepackage{graphicx}
\usepackage{amsmath,amssymb}
\usepackage{url}
\usepackage{booktabs}
\usepackage{caption}
\usepackage{subcaption}
\usepackage{multirow}
\usepackage{hyperref}
\usepackage{tikz}
\usepackage{geometry}
\usepackage{amssymb}
\usepackage{pifont}
\usepackage{comment}
\newcommand{\cmark}{\checkmark}
\newcommand{\xmark}{\ding{55}}
\usetikzlibrary{shapes.geometric, arrows, positioning, fit}

\tikzstyle{mainbox} = [rectangle, rounded corners, minimum width=4cm, minimum height=1.5cm, text centered, draw=black, fill=white, text width=4cm]
\tikzstyle{excludebox} = [rectangle, rounded corners, minimum width=3cm, minimum height=1.5cm, text centered, draw=red, fill=red!10, text width=3cm]
\tikzstyle{arrow} = [thick,->,>=stealth]
\tikzstyle{redarrow} = [thick,->,>=stealth,red]

\begin{document}

\title{A Comprehensive Review of Denial of Wallet Attacks in Serverless Architectures}

\author{\IEEEauthorblockN{Mark Dorsett, Scott Mann, Jabed Chowdhury, Abdun Mahmood}\\
\IEEEauthorblockA{\textit{La Trobe University} \\
\{M.Dorsett, S.Mann, J.Chowdhury, AMahmood\}@latrobe.edu.au}}

\maketitle

\begin{abstract}
The Denial of Wallet (DoW) attack poses a unique and growing threat to serverless architectures that rely on Function-as-a-Service (FaaS) models, exploiting the cost structure of pay-as-you-go billing to financially burden application owners. Unlike traditional Denial of Service (DoS) attacks, which aim to exhaust resources and disrupt service availability, DoW attacks focus on escalating costs without impacting service operation. This review traces the evolution of DoW research, from initial awareness and attack classification to advancements in detection and mitigation strategies. Key developments include the categorisation of attack types—such as Blast DDoW, Continual Inconspicuous DDoW, and Background Chained DDoW—and the creation of simulation tools like DoWTS, which enable safe experimentation and data generation. Recent advancements highlight machine learning approaches, including systems like Gringotts and DoWNet, which leverage deep learning and anomaly detection to identify malicious traffic patterns. Although substantial progress has been made, challenges persist, notably the lack of real-world data and the need for adaptive billing models. This is the first comprehensive literature review dedicated strictly to Denial of Wallet attacks, providing an in-depth analysis of their financial impacts, attack techniques, mitigation strategies, and detection mechanisms within serverless computing. The paper also presents the first detailed examination of simulation and data generation tools used for DoW research, addressing a critical gap in existing cybersecurity literature. By synthesising these key areas, this study serves as a foundational resource for future research and industry efforts in securing pay-as-you-go cloud environments. 
\end{abstract}


\section{Introduction}

The rise of cloud computing, specifically serverless architectures, has significantly reshaped the deployment and economic structures of cloud-based applications. Serverless platforms, operating on a Function-as-a-Service (FaaS) model, empower developers to deploy code as discrete functions that execute in response to specific triggers, eliminating the need for constant infrastructure management. This model is particularly attractive for businesses due to its cost efficiency, as charges are based strictly on function execution time, translating to a flexible ``pay-as-you-go'' billing approach.

However, the pay-as-you-go model introduces unique security and financial vulnerabilities, one of which is the Denial of Wallet (DoW) attack \cite{kelly2021denial}. Unlike traditional Denial of Service (DoS) attacks that overwhelm system resources to disrupt availability, DoW attacks aim to exploit the financial structure of serverless environments. By maliciously triggering function invocations all at once, attackers drive up the victim’s usage costs, potentially leading to substantial financial burdens without affecting service availability. This economic angle of exploitation, distinct from resource-exhausting DoS attacks, has made DoW an area of growing concern within the cybersecurity community \cite{kelly2021denial, ortega2024generation, mileski2022distributed}.

DoW’s impact on the cybersecurity scene has become increasingly significant as more companies adopt serverless models. In particular, industries with fluctuating and high traffic demands, such as e-commerce, fintech, and content streaming, are at heightened risk. In these environments, DoW attacks can go undetected due to legitimate variations in usage, allowing attackers to steadily drain financial resources. As serverless architectures continue to expand, understanding and mitigating the risk of DoW attacks has become important, particularly for business sustainability. The following sections examine the progression of research on DoW attacks, categorising progress across different themes and highlighting pivotal advancements. The following are our contributions to the literature: first literature review on DoW attacks, categorisation and taxonomy of DoW attack techniques, analysis of financial impact studies of denial of wallet attacks, comparison of common denial of wallet attack types, comparison of denial of wallet mitigation strategies, ML and AI Solutions for DoW attack mitigation, recommended ML/AI Solutions per DoW attack type, discussion, and challenges and future direction.

\section{Comparison of Existing Literature Reviews}

An in-depth analysis of nineteen prominent survey papers on Distributed Denial of Service (DDoS) attacks from 2013 to 2025 reveals several compelling trends and notable research gaps. These patterns offer insight into the evolution of cybersecurity research and position the present study as a significant and timely contribution that addresses all critical and previously underexplored areas.

A prominent trend seen in the literature is the increased reliance on machine learning (ML) and artificial intelligence (AI) techniques for DDoS detection. 

While only a few research works \cite{yan2016software} addressed ML-based detection prior to 2020, this topic has gained considerable traction in recent years.

From 2020 onward, six of the ten most recent papers incorporated ML/AI approaches, reflecting a shift towards adaptive, data-driven security systems that can recognise evolving and obfuscated attack signatures. This surge underscores the academic community’s response to the growing sophistication of modern DDoS vectors, especially within IoT and software-defined network (SDN) environments.

In contrast, Economic Denial of Sustainability (EDoS), a form of denial attack that exploits the cost models of cloud infrastructure, remains critically underexplored. Of the 19 papers reviewed, only Alashhab et al. \cite{alashhab2021impact} explicitly addresses EDoS in depth, despite its growing relevance in cloud-native and serverless architectures. Similarly, financial impact analysis is rarely explored. This is surprising given the fundamental shift in cloud computing towards metered, usage-based billing models that are inherently vulnerable to financially motivated attacks. The absence of financial analysis in most DDoS literature represents a critical oversight in understanding the full scope of modern denial threats.

Another important observation pertains to the growing use of simulation environments and datasets in research published after 2022. Before this period, papers primarily provided conceptual or taxonomy-based reviews. However, recent studies \cite{balarezo2022survey, ma2025survey} integrate empirical validation through simulated testbeds or publicly available datasets, marking a methodological advancement towards reproducibility and practical benchmarking in security research.

In contrast to these selective trends, mitigation and defence strategies are universally discussed across all studies, regardless of publication year. This consistency highlights the foundational nature of defensive techniques in DDoS literature and reflects the field’s emphasis on applied solutions. Researchers have consistently focused on identifying, categorising, and evaluating mitigation mechanisms, from traffic filtering and source authentication to rate limiting and AI-driven anomaly detection.

A striking and consistent gap across the literature is the absence of discussion on serverless computing and Function-as-a-Service (FaaS) in depth. Despite the growing prevalence of serverless architectures and their unique security implications, none of the 19 papers analysed include a dedicated focus on these examples. Given their reliance on auto-scaling and pay-as-you-go billing, serverless platforms are uniquely susceptible to denial attacks. This silence across nearly a decade of survey literature points to a critical area in need of academic attention.

In contrast to these gaps, this paper addresses all ten core topics: Denial Attack, Economic Denial of Sustainability (EDoS), Cloud Computing, Serverless Architecture, FaaS, Financial Impact Analysis, Attack Techniques, Simulation and Data Generation, Mitigation and Defence Strategies, and Machine Learning and AI Detection. By bridging the disconnects in current literature, this research offers a comprehensive and contemporary understanding of denial-based attacks and their implications in evolving cloud environments.

\begin{figure}
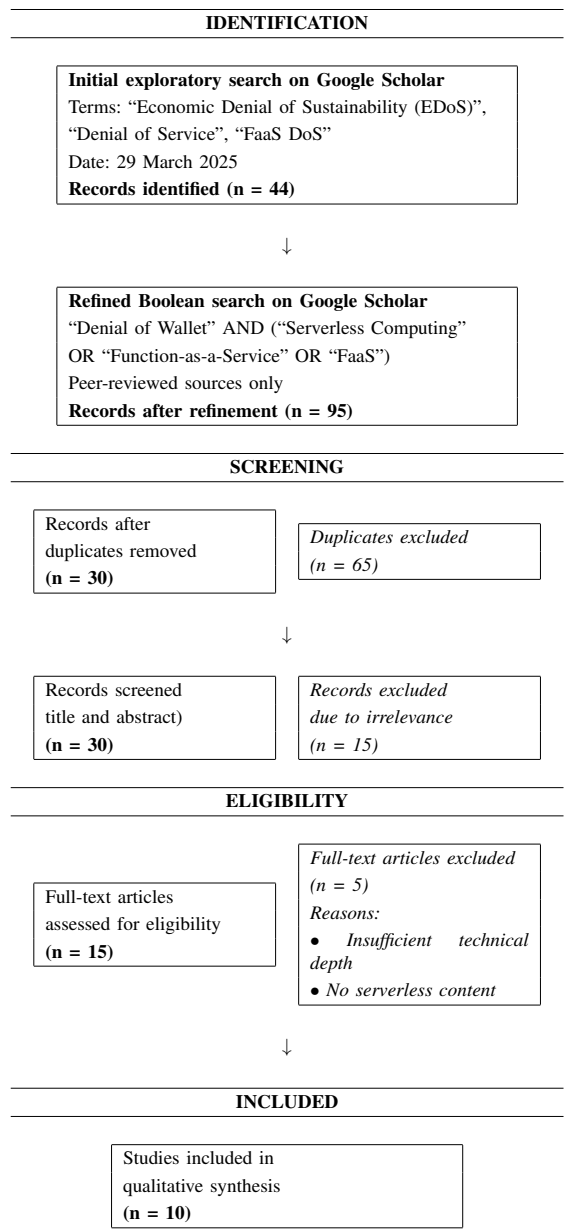

\centering
\renewcommand{\arraystretch}{1.2}
\resizebox{0.95\linewidth}{!}{%
\begin{tabular}{c}
\hline
\textbf{IDENTIFICATION} \\
\hline
\\
\begin{tabular}{|p{8cm}|}
\hline
\textbf{Initial exploratory search on Google Scholar} \\
Terms: ``Economic Denial of Sustainability (EDoS)'', \\
``Denial of Service'', ``FaaS DoS'' \\
Date: 29 March 2025 \\
\textbf{Records identified (n = 44)} \\
\hline
\end{tabular} \\
\\
$\downarrow$ \\
\\
\begin{tabular}{|p{8cm}|}
\hline
\textbf{Refined Boolean search on Google Scholar} \\
``Denial of Wallet'' AND (``Serverless Computing'' \\
OR ``Function-as-a-Service'' OR ``FaaS'') \\
Peer-reviewed sources only \\
\textbf{Records after refinement (n = 95)} \\
\hline
\end{tabular} \\
\\
\hline
\textbf{SCREENING} \\
\hline
\\
\begin{tabular}{cc}
\begin{tabular}{|p{4cm}|}
\hline
Records after \\
duplicates removed \\
\textbf{(n = 30)} \\
\hline
\end{tabular}
&
\begin{tabular}{|p{4cm}|}
\hline
\textit{Duplicates excluded} \\
\textit{{(n = 65)}} \\
\hline
\end{tabular}
\end{tabular} \\
\\
$\downarrow$ \\
\\
\begin{tabular}{cc}
\begin{tabular}{|p{4cm}|}
\hline
Records screened \\
title and abstract) \\
\textbf{(n = 30)} \\
\hline
\end{tabular}
&
\begin{tabular}{|p{4cm}|}
\hline
\textit{Records excluded} \\
\textit{due to irrelevance} \\
\textit{{(n = 15)}} \\
\hline
\end{tabular}
\end{tabular} \\
\\
\hline
\textbf{ELIGIBILITY} \\
\hline
\\
\begin{tabular}{cc}
\begin{tabular}{|p{4cm}|}
\hline
Full-text articles \\
assessed for eligibility \\
\textbf{(n = 15)} \\
\hline
\end{tabular}
&
\begin{tabular}{|p{4cm}|}
\hline
\textit{Full-text articles excluded} \\
\textit{(n = 5)} \\
\textit{Reasons:} \\
\textit{$\bullet$ Insufficient technical depth} \\
\textit{$\bullet$ No serverless content} \\
\hline
\end{tabular}
\end{tabular} \\
\\
$\downarrow$ \\
\\
\hline
\textbf{INCLUDED} \\
\hline
\\
\begin{tabular}{|p{6cm}|}
\hline
Studies included in \\
qualitative synthesis \\
\textbf{(n = 10) }\\
\hline
\end{tabular} \\
\\
\end{tabular}
}
\caption{Systematic literature review process for identifying studies on Denial of Wallet (DoW) attacks in serverless computing environments.}
\label{fig:prisma_flow}
\end{figure}

%original table, do not delete
\begin{comment}
\end{comment}

\begin{table*}[t]
\centering
\caption{Review of Related Surveys and Reviews on Denial Attacks}
\label{tab:literature_comparison}
\renewcommand{\arraystretch}{1.2}
\resizebox{0.999\textwidth}{!}{%
\begin{tabular}{|l|c|c|c|c|c|c|c|c|c|c|c|}
\hline
\textbf{Paper} & \textbf{Denial Attack} & \textbf{EDoS} & \textbf{Cloud} & \textbf{Serverless} & \textbf{FaaS} & \textbf{Fin. Impact} & \textbf{Attack Tech.} & \textbf{Sim. \& Data} & \textbf{Mitigation} & \textbf{ML/AI} & \textbf{Year} \\
\hline
Zargar et al. \cite{zargar2013} & \cmark & \xmark & \cmark & \xmark & \xmark & \xmark & \cmark & \xmark & \cmark & \xmark & 2013 \\
Yan et al. \cite{yan2016software} & \cmark & \xmark & \cmark & \xmark & \xmark & \xmark & \cmark & \xmark & \cmark & \cmark & 2016 \\
Mallikarjunan et al. \cite{mallikarjunan2016survey} & \cmark & \xmark & \cmark & \xmark & \xmark & \xmark & \cmark & \xmark & \cmark & \xmark & 2016 \\
Dong et al. \cite{dong2019survey} & \cmark & \xmark & \cmark & \xmark & \xmark & \xmark & \cmark & \xmark & \cmark & \cmark & 2019 \\
Odusami et al. \cite{odusami2020survey} & \cmark & \xmark & \cmark & \xmark & \xmark & \xmark & \cmark & \xmark & \cmark & \cmark & 2020 \\
Tandon \cite{tandon2020survey} & \cmark & \xmark & \cmark & \xmark & \xmark & \xmark & \cmark & \xmark & \cmark & \xmark & 2020 \\
Salim et al. \cite{salim2020survey} & \cmark & \xmark & \cmark & \xmark & \xmark & \xmark & \cmark & \xmark & \cmark & \cmark & 2020 \\
Khader \& Eleyan \cite{khader2021survey} & \cmark & \xmark & \cmark & \xmark & \xmark & \xmark & \cmark & \xmark & \cmark & \xmark & 2021 \\
Eliyan \& Di Pietro \cite{eliyan2021survey} & \cmark & \xmark & \cmark & \xmark & \xmark & \xmark & \cmark & \xmark & \cmark & \xmark & 2021 \\
Al-Hadhrami \cite{alhadhrami2021survey} & \cmark & \xmark & \cmark & \xmark & \xmark & \xmark & \cmark & \xmark & \cmark & \cmark & 2021 \\
Dalmazo et al. \cite{dalmazo2021survey} & \cmark & \xmark & \cmark & \xmark & \xmark & \xmark & \cmark & \xmark & \cmark & \cmark & 2021 \\
Shah et al. \cite{shah2022survey} & \cmark & \xmark & \cmark & \xmark & \xmark & \xmark & \cmark & \xmark & \cmark & \xmark & 2022 \\
Alashhab et al. \cite{alashhab2021impact} & \cmark & \cmark & \cmark & \xmark & \xmark & \xmark & \cmark & \xmark & \cmark & \cmark & 2022 \\
Balarezo et al. \cite{balarezo2022survey} & \cmark & \xmark & \cmark & \xmark & \xmark & \xmark & \cmark & \cmark & \cmark & \cmark & 2022 \\
Alhijawi et al. \cite{alhijawi2022survey} & \cmark & \xmark & \cmark & \xmark & \xmark & \xmark & \cmark & \cmark & \cmark & \cmark & 2022 \\
De Neira et al. \cite{deneira2023survey} & \cmark & \xmark & \cmark & \xmark & \xmark & \xmark & \cmark & \cmark & \cmark & \cmark & 2023 \\
Ali et al. \cite{ali2023mlsurvey} & \cmark & \xmark & \cmark & \xmark & \xmark & \xmark & \cmark & \xmark & \cmark & \cmark & 2023 \\
Ma et al. \cite{ma2025survey} & \cmark & \xmark & \cmark & \xmark & \xmark & \xmark & \cmark & \cmark & \cmark & \cmark & 2025 \\
\textbf{This Work} & \cmark & \cmark & \cmark & \cmark & \cmark & \cmark & \cmark & \cmark & \cmark & \cmark & 2025 \\
\hline
\end{tabular}
}
\end{table*}

\section{Systematic Literature Methodology}

This study employed a systematic search methodology aligned with the PRISMA framework \cite{page2021prisma} to identify relevant academic literature pertaining to Denial of Wallet (DoW) attacks in serverless computing environments. The objective was to capture high-quality, peer-reviewed publications that substantively address the technical, architectural, or financial dimensions of DoW threats.

An initial exploratory search was conducted using Google Scholar with the following broad terms: ``Economic Denial of Sustainability (EDoS),'' ``Denial of Service,'' and ``FaaS DoS.'' This search yielded 44 unique records. However, many of these were subsequently excluded due to their lack of specificity to the Denial of Wallet concept.

Following iterative refinement and preliminary screening, it was determined that the term ``Denial of Wallet'' was the most appropriate and accurate descriptor for the attack class under investigation. Nevertheless, a search using only this term returned a high number of irrelevant results, many of which included passing mentions of DoW without meaningful technical engagement.

To enhance the relevance of results, a Boolean-enhanced query was adopted:

\begin{quote}
\texttt{"Denial of Wallet" AND ("Serverless Computing" OR "Function-as-a-Service" OR "FaaS")}
\end{quote}

This modified search string significantly improved result precision by targeting articles that linked DoW discussions explicitly to serverless architectures. The search was performed on Google Scholar, given its extensive academic coverage, on 29 March 2025. The term ``Denial of Wallet'' returned approximately 95 results on Google Scholar and 24,500 results via Google Search. However, web pages, blog posts, and other non-peer-reviewed sources were excluded from analysis.

We have therefore used the following inclusion and exclusion criteria in this survey paper:

Inclusion criteria required that articles:

\begin{itemize}
    \item Be published in peer-reviewed journals or conference proceedings.
    \item Be written in English.
    \item Provide substantive research of Denial of Wallet attacks, particularly in the context of serverless, Function-as-a-Service (FaaS), or cloud-native architectures.
\end{itemize}

Articles were excluded if they:

\begin{itemize}
    \item Lacked technical analysis or experimental data on DoW.
    \item Only mentioned DoW in a cursory manner.
    \item Were grey literature, such as blogs or non-peer-reviewed whitepapers.
    \item Focused on Distributed Denial of Service rather than DoW.
\end{itemize}

After removal of duplicates and the application of inclusion and exclusion criteria, 30 records proceeded to the title and abstract screening stage. Of these, 15 were excluded due to irrelevance. The remaining 15 full-text articles were retrieved and assessed in detail. A further five articles were excluded at this stage due to insufficient technical depth or absence of serverless-related content.

Ultimately, 10 articles were included in the final review. These publications constitute the foundational literature base for the synthesis and analysis presented in subsequent sections.  We understand there is a significant lack of studies covered in this literature review; this is because the topic is infant in nature, state-of-the-art and there is a lack of awareness relating to denial attacks, which is supported by our industry expert survey on denial attack awareness and financial impact \cite{dorsett2025ddos}.\\

The remainder of this paper is structured as follows: Section IV provides background on denial-based threats, expressing DoW to DoS, DDoS, and EDoS. Section V highlights the awareness and early identification of DoW. Section VI presents our study and real-world data on the impacts of denial of wallet attacks. Section VII classifies denial of wallet attack methods. Section VIII highlights existing DoW simulations and data sets. Section IX categorises mitigation strategies. Section X evaluates AI and ML-based detection techniques. Section XI discusses limitations, and Section XII concludes with future research directions.

\section{Background}

Denial of Wallet (DoW) attacks exploit the pay-as-you-go billing model in serverless cloud environments to inflict financial damage without disrupting availability. Unlike Denial of Service (DoS) attacks, DoW abuses auto-scaling and Function-as-a-Service (FaaS) platforms like AWS Lambda to trigger excessive function invocations. The abstraction and limited visibility of cloud infrastructure further complicate detection. As serverless systems scale on demand, they become ideal targets for cost-based attacks, highlighting the need for financially-aware cybersecurity strategies in modern cloud-native architectures.

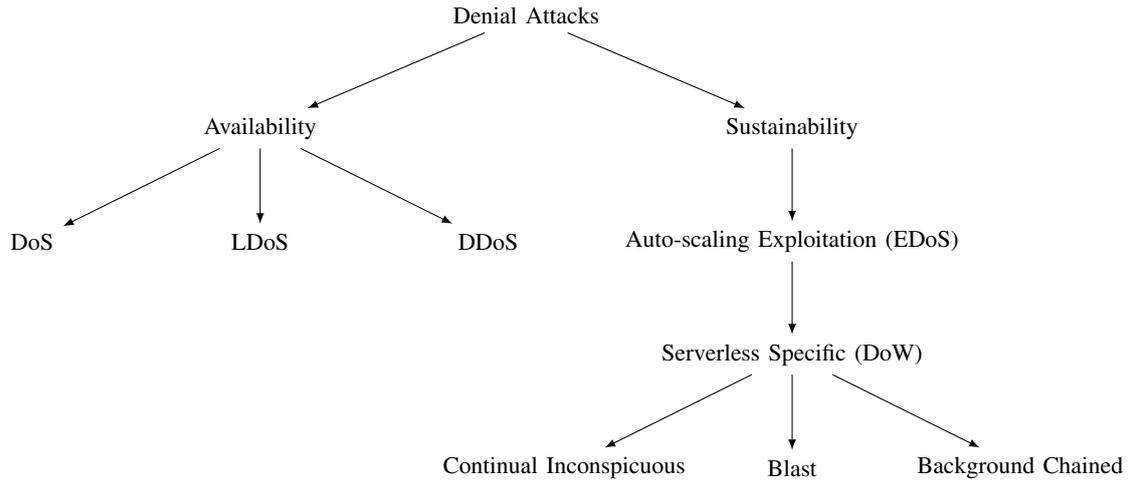
\begin{figure*}[t]
\centering
\scalebox{1}{ % Adjust this for overall size scaling if needed
\begin{tikzpicture}[
  sibling distance=30pt,
  every node/.style = {font=\small},
  level distance=1.5cm,
  level 1/.style={sibling distance=70mm},
  level 2/.style={sibling distance=30mm},
  edge from parent/.style={draw,-latex}
]
% Root node
\node {Denial Attacks}
  child { node {Availability}
    child { node {DoS} }
    child { node {LDoS} }
    child { node {DDoS} }
  }
  child { node {Sustainability}
    child { node {Auto-scaling Exploitation (EDoS)}
      child { node {Serverless Specific (DoW)}
        child { node {Continual Inconspicuous} }
        child { node {Blast} }
        child { node {Background Chained} }
      }
    }
  };
\end{tikzpicture}
} % end scalebox
\caption{Comparative Taxonomy of Denial Attacks: Availability vs. Sustainability}
\label{fig:taxonomy}
\end{figure*}

Economic Denial of Sustainability (EDoS) is a form of economic attack targeting cloud-native applications. Unlike traditional denial-based threats that seek to disrupt service availability, DoW attacks exploit the cost structures of cloud environments by deliberately inflating usage and triggering billable events. These attacks aim to financially exhaust victims by exploiting elastic scaling features and metered resource models, often without disrupting performance, making them difficult to detect using conventional threat monitoring approaches.

Denial of Service (DoS) attacks are designed to overwhelm systems, networks, or applications to render services unavailable to legitimate users. These attacks typically rely on flooding targets with excessive traffic or requests, exhausting computational or bandwidth resources. While DoS impacts availability, it does not necessarily induce direct financial damage. In contrast, DoW attacks adopt a more covert, cost-focused strategy, demonstrating a paradigm shift in the motives and mechanisms of denial-based threats.

Serverless architecture abstracts infrastructure management from developers, enabling code execution without provisioning or maintaining servers. In this paradigm, cloud providers dynamically manage resources, scaling functions in response to demand. While this approach offers significant benefits in terms of scalability and cost-efficiency, its on-demand nature also exposes applications to new threat vectors, such as DoW attacks, which exploit the automatic invocation and billing mechanics inherent in serverless platforms.

Function-as-a-Service (FaaS) is a core component of serverless computing that enables developers to deploy discrete, stateless functions triggered by predefined events. These functions execute in ephemeral containers and scale automatically based on event frequency. While FaaS simplifies deployment and optimises resource usage, it also introduces new risks, as its fine-grained billing model and seamless scalability can be exploited by adversaries aiming to inflate costs through excessive or chained function invocations.

The pay-as-you-go billing model underpins most cloud computing services, charging users only for the compute, storage, or networking resources they consume. This model supports cost efficiency, flexibility, and scalability. However, it also introduces financial exposure, as attackers can manipulate service usage patterns to trigger cost escalation without compromising system performance. This inherent vulnerability is central to the feasibility and impact of Denial of Wallet attacks in serverless environments.

Cloud technology provides scalable, on-demand access to computing resources over the internet, supporting a wide range of services from infrastructure to applications. Its abstraction of physical hardware, combined with elastic scalability, has revolutionised software deployment models. However, the distributed and managed nature of cloud services reduces direct visibility and control, challenging traditional security approaches and creating fertile ground for financially driven attacks like DoW.

Auto-scaling is a key feature of cloud platforms that automatically adjusts resource allocation based on current demand. It enables high availability and performance optimisation by launching or terminating function instances in real-time. While beneficial for handling variable workloads, auto-scaling can be weaponised in DoW attacks, as attackers can trigger high-cost scaling events, causing substantial financial impact without degrading service responsiveness or availability.

\section{Awareness and Early Identification of DoW}

The evolution of cloud computing, particularly the advent of serverless architectures, has inadvertently introduced new vectors for economically motivated denial attacks. Among these, Denial of Wallet (DoW) represents a modern variant of resource consumption attacks designed to exploit cloud billing models by incurring persistent or unsustainable operational costs to the victim. Conceptually rooted in the earlier Economic Denial of Sustainability (EDoS) attack, DoW differs by targeting billing thresholds rather than service unavailability, rendering financial insolvency the primary attack outcome rather than degraded system performance.

The term Economic Denial of Sustainability (EDoS) was first introduced by Hoffman in 2008 \cite{hoffman2008edos}, identifying a scenario wherein malicious actors exploit auto-scaling features of cloud systems to artificially inflate resource usage and, consequently, operational costs. Unlike traditional Distributed Denial of Service (DDoS) attacks, which aim to exhaust bandwidth or processing power, EDoS attacks exploit the ``pay-as-you-go'' pricing model of cloud computing, creating long-term financial degradation without necessarily disrupting availability.

Building upon this, Denial of Wallet (DoW) emerged as a distinct attack classification in the mid-2010s, coinciding with the mass adoption of Function-as-a-Service (FaaS) platforms. The 2014 introduction of AWS Lambda \cite{ortega2024generation}, followed by Azure Functions in 2016 and Google Cloud Functions in 2017, offered unprecedented scalability with billing based on precise execution time and number of invocations. While these models provided operational flexibility, they also exposed a critical vulnerability: cost proportionality. By exploiting these billing triggers, often through low-rate, inconspicuous API calls, attackers could drive up usage-based costs without triggering conventional DDoS alarms.

While early discussions around DoW remained largely anecdotal or theoretical, formal academic recognition began to solidify by 2021, with works such as Kelly et al. \cite{kelly2021denial} and Mileski and Mihajloska \cite{mileski2022distributed} investigating the impact of denial of wallet attacks in elastic cloud environments. Industry awareness of DoW attacks has grown in parallel with the rise of serverless adoption. Security bulletins by cloud providers such as AWS and Google began referencing cost-based attack vectors as early as 2018, albeit without standardized terminology. Concurrently, cybersecurity researchers proposed taxonomies that classified DoW as a distinct subclass of cloud-specific threats, alongside EDoS and Resource Exhaustion Attacks.

Awareness initiatives have also expanded into practitioner communities. For instance, the Cloud Security Alliance (CSA) has included EDoS and cost-exhaustion risks in its threat modelling guidance since 2019, emphasising the importance of monitoring financial metrics in tandem with performance indicators.

Despite growing awareness, underreporting remains a critical issue. Many DoW attacks are difficult to distinguish from legitimate traffic patterns, especially when conducted at low rates over extended periods—a tactic often referred to as Continual Inconspicuous DoW. This challenge has created a lag in widespread adoption of countermeasures, particularly among small and medium enterprises with limited cloud monitoring capabilities.

Although detailed reports of real-world DoW attacks are scarce due to non-disclosure by cloud providers and enterprises, controlled academic simulations have effectively demonstrated the feasibility and economic impact of such attacks. For instance, in simulated environments, researchers have shown that an attacker could incur hundreds or thousands of dollars in additional billing by triggering Lambda functions using minimally invasive API calls every few seconds for a month \cite{mileski2022distributed, kelly2023dowts, kelly2024downet}.

\section{Real-world Financial Impact of Denial of Wallet}

Empirical data on the financial impact of denial-of-service attacks, and more specifically Denial of Wallet (DoW) attacks, remains exceptionally scarce in both academic literature and industry reporting. This scarcity is likely the result of two compounding factors. First, organisations often lack awareness of financially motivated cloud-native attack vectors such as DoW, and thus fail to detect or classify them accurately. Second, even when financial damage is recognised, organisations are frequently unwilling to disclose the extent of monetary loss, driven by a desire to protect shareholder value, brand reputation, and consumer confidence. Consequently, there exists a systemic data vacuum that undermines the development of effective mitigation strategies and inhibits sector-wide benchmarking. This underreporting is particularly problematic given the increasing prevalence of Function-as-a-Service (FaaS) models, which are vulnerable to cost-exhaustion attacks that do not necessarily disrupt service availability.

As demonstrated in a 2024 post by a software engineer \cite{stack2024aws}, it was shown that an attacker can exploit such vulnerabilities by sending a high volume of rejected requests to an AWS S3 bucket, thereby inflating the victim’s bill significantly—up to USD\$1,300 a day—without breaching the system or halting service. These emerging attack patterns reveal the financial fragility of modern cloud architectures, and the urgency of improved detection and reporting mechanisms for economically driven attacks.

% REPLACING THE BELOW SECTION WITH THE SURVEY STUDY WE CONDUCTED DIRECTLY IN HERE. DO NOT REMOVE THE BELOW COMMENT.

Due to the lack of exposure and real-world data on DoW, we conducted the following study to gather insights on industry awareness relating to denial attacks \cite{dorsett2025ddos}. 
Our industry study offers one of the few existing and available empirical snapshots of organisational awareness and preparedness against denial attacks, including DoW attacks.

Our study, administered to cybersecurity and network professionals across multiple Australian sectors including ICT, Law and Government, Education, and Finance, revealed a significant knowledge gap. Although 100\% of participants were aware of traditional DDoS attacks, only 13.6\% reported familiarity with DoW attacks, and not a single respondent could confirm whether their organisation had ever been a victim. This highlights a critical blind spot in the industry’s ability to monitor or respond to financially motivated FaaS abuse.

The knowledge gap is matched by a shortfall in financial risk governance. While 86\% of surveyed organisations reported implementing formal cybersecurity risk assessments using frameworks such as NIST, ISO 27001, Essential 8, and PCI-DSS, these assessments did not extend to the financial implications of FaaS deployments in most cases. In fact, 64\% of organisations that actively deploy serverless functions reported that they do not assess the financial risks associated with such services. This is particularly troubling given that the very nature of a DoW attack is to exploit billing structures in FaaS environments to create economic harm. The findings indicate a clear disconnect between cybersecurity maturity models and financial risk modelling, exposing organisations to invisible cost-based attacks.

Among the 22.7\% of respondents who claimed to assess financial risks tied to FaaS usage, no consistent methodologies or tools were identified. Responses included internal risk checklists, informal reviews of cost implications, and considerations of factors such as resource consumption, regulatory compliance, and latency. However, there was no evidence of standardised frameworks or predictive modelling approaches that could offer reliable protection against DoW threats. Moreover, less than 15\% of participants were able to quantify the financial consequences of their most recent DDoS incident, underscoring the broader industry challenge of economic impact analysis following cybersecurity breaches.

Notably, the survey data suggests that organisational size plays a role in risk visibility. Enterprises with 5,000 to 49,999 employees were more likely to conduct FaaS financial risk assessments than smaller firms (150–999 employees), likely due to differences in available resources, maturity of IT governance, and budget allocations. However, even among larger organisations, the lack of post-incident cost forensics was evident. Many respondents cited difficulties in tracing economic fallout due to insufficient monitoring, lack of integration between financial and security teams, and the absence of cloud-native billing anomaly detection systems.

Taken together, these findings indicate that Denial of Wallet attacks represent an under-acknowledged and unquantified threat in contemporary cloud environments. Despite the increasing adoption of cloud-native services and risk assessment protocols, the failure to integrate financial modelling into security governance leaves organisations exposed to cost-exhaustive attack vectors that may go undetected until billing anomalies arise. This silent mode of disruption contrasts sharply with conventional DDoS attacks, which are loud, overt, and easier to attribute.

To address this issue, we suggest future research and industry practice must converge on four priorities:
\begin{enumerate}
    \item Developing standardised financial risk assessment models for FaaS deployments;
    \item Expanding awareness campaigns on DoW and EDoS threats within enterprise IT and finance teams;
    \item Integrating cloud billing telemetry and usage anomaly detection into cybersecurity monitoring frameworks;
    \item Establishing a standardised taxonomy for each denial attack, allowing professionals to distinguish which denial attack they’re defending against.
\end{enumerate}

Only through such multidisciplinary approaches can the financial integrity of cloud-based operations be safeguarded against the next generation of denial-of-service attacks.

\section{Classification and Attack Types}

The next advancement in DoW research occurred as scholars began classifying DoW attacks to better understand their mechanics and distinguish them from other forms of financial threats, such as EDoS. By categorising DoW attacks into distinct types, researchers developed a structured framework for analysing the diverse methodologies employed in these attacks. Mileski and Mihajloska \cite{mileski2022distributed} define three distinct types of DoW attacks: \textit{Blast DDoW}, \textit{Continual Inconspicuous DDoW}, and \textit{Background Chained DDoW}.

\textbf{Blast DDoW:} This high-intensity attack generates a massive number of function invocations within a brief period, maximising the victim’s billing costs by exploiting the serverless platform’s auto-scaling capability. Given that serverless applications are designed to scale seamlessly, they can absorb large traffic volumes without downtime, effectively insulating users from the disruption seen in DoS attacks. However, in the context of a Blast DDoW, the financial impact is immediate and substantial. Research has shown that such attacks can target specific function endpoints, maximising invocation rates to multiply costs in a short time. Studies indicate that major cloud platforms may have some rate-limiting mechanisms, but these often fall short in preventing the rapid scaling costs typical of Blast DDoW \cite{mileski2022distributed, kelly2024downet}.

\textbf{Continual Inconspicuous DDoW:} Operating in a less obvious manner than Blast DDoW, this method involves triggering a steady, low-level stream of function invocations over an extended period. By maintaining a prolonged but manageable rate, attackers remain under the radar of conventional detection tools. Studies reveal that this ``slowburn'' tactic is particularly effective in environments with consistent background traffic, as attackers can blend malicious requests with legitimate traffic to evade detection. Unlike high-volume attacks, Continual Inconspicuous DDoW presents unique challenges for detection systems, as it operates within normal traffic thresholds yet gradually incurs significant costs \cite{mileski2022distributed}.

\textbf{Background Chained DDoW:} This complex form of attack leverages the interconnected nature of serverless functions to amplify its impact. By initiating one serverless function that, in turn, triggers additional services, attackers create a cascade of invocations that can multiply costs exponentially. Background Chained DDoW takes advantage of automated workflows and microservice architectures, where interconnected functions are common. Studies have indicated that this attack type is particularly effective in environments where functions are interdependent, making it challenging to isolate the malicious trigger from legitimate workflows \cite{kelly2024downet}.

Each classification highlights a different approach to achieving the same goal: the financial exhaustion of the victim. By categorising DoW attacks into these distinct types, researchers have underscored the need for diverse detection and mitigation strategies tailored to each attack’s specific characteristics.

\begin{table*}[t]
\centering
\caption{Comparison of Denial of Wallet Attack Types}
\label{tab:attack_types}
\renewcommand{\arraystretch}{1.2}
\resizebox{0.999\textwidth}{!}{%
\begin{tabular}{|p{3cm}|p{3.2cm}|p{2.4cm}|p{2.2cm}|p{3.2cm}|p{3.5cm}|}
\hline
\textbf{Attack Type} & \textbf{Attack Strategy} & \textbf{Goal} & \textbf{Visibility} & \textbf{Detection Challenge} & \textbf{Deployment Context} \\
\hline
Blast DDoW & High-volume function invocations in a short time & Immediate cost spike through rapid auto-scaling & High – easily noticeable due to traffic spikes & Rate-limiting may slow but not prevent cost escalation & Auto-scaling and function concurrency \\
\hline
Continual Inconspicuous DDoW & Low-level, continuous invocation stream over long periods & Stealthy cost accumulation over time & Low – blends with normal traffic patterns & Difficult to distinguish from legitimate background activity & Background activity and consistent low-volume workloads \\
\hline
Background Chained DDoW & Trigger chains of interconnected functions for exponential invocation & Exponential cost multiplication through cascading function calls & Medium – obscured by function dependencies and workflows & Hard to trace origin due to complex service chaining & Function orchestration and microservice chaining \\
\hline
\end{tabular}
}
\end{table*}

\section{Simulation and Data Generation}

The financial risks associated with testing DoW attacks on live serverless platforms have necessitated the development of safe, controlled environments for research and experimentation. One of the most significant advancements in this area has been the development of the Denial-of-Wallet Test Simulator (DoWTS) \cite{kelly2023dowts}. This tool enables researchers to replicate DoW scenarios in a simulated serverless environment, generating synthetic data that closely mimics real-world attack patterns.

DoWTS offers a multifaceted testing ground for researchers, providing not only the ability to simulate high-frequency attacks but also to experiment with low-visibility, prolonged attacks. By allowing researchers to observe attack impacts in a financially neutral environment, DoWTS has become an invaluable resource for both studying the efficacy of detection systems and developing new mitigation techniques. Importantly, DoWTS generates labelled datasets, a critical resource for training machine learning models that require vast amounts of structured data to identify malicious patterns accurately \cite{kelly2023dowts, ortega2024generation}.

The use of synthetic data has enabled researchers to overcome a persistent challenge in DoW research: the scarcity of real-world data. Organisations are often reluctant to disclose DoW incidents due to financial or reputational risks, leading to a gap in publicly available data. DoWTS and similar synthetic data generation tools bridge this gap, providing the research community with the data necessary to simulate real-world scenarios and refine detection methodologies. Despite this progress, reliance on synthetic data introduces limitations, as real-world traffic variability and noise can be difficult to replicate accurately. Nevertheless, DoWTS remains a foundational tool for advancing DoW research and developing robust, scalable defence mechanisms.

\section{Mitigation and Defence Strategies}

Mitigating DoW attacks requires a nuanced approach, balancing between preventing financial exploitation and maintaining legitimate access for users. Research on DoW mitigation strategies has led to several innovative solutions, each aimed at countering specific vulnerabilities inherent in serverless environments. The key methods include API rate limiting, execution time limits, cost management alerts, and adaptive billing caps.

\textbf{API Rate Limiting and Throttling:} By restricting the number of requests an API endpoint can handle within a set period, rate limiting prevents rapid scaling of function invocations, which is particularly effective against Blast DDoW attacks. Throttling adds an extra layer of control, slowing down function execution when certain thresholds are met. However, this approach has drawbacks, as it can also limit legitimate traffic during high-demand periods, affecting service quality. For instance, industries like ecommerce may experience increased traffic during peak times, which could inadvertently trigger rate-limiting measures and impact customer experience \cite{kelly2024downet, kelly2022poster}.

\textbf{Execution Time Limits:} Setting strict time limits for function executions can mitigate attacks that rely on prolonged function invocations to inflate costs. This approach is especially useful for services with well-defined processing requirements. However, as studies indicate, execution limits can be challenging to implement in functions that require variable processing times, such as data analysis applications. Balancing time limits to avoid unnecessary interruptions of legitimate processes remains a challenge, requiring careful calibration based on specific service needs \cite{mileski2022distributed}.

\textbf{Cost Management Alerts:} Adaptive cost alerts serve as an early warning system, notifying administrators when function usage exceeds predefined financial thresholds. Cloud providers have started to incorporate such features, enabling application owners to monitor costs in real-time. Studies show that these alerts are particularly effective for detecting low-intensity attacks that may otherwise go unnoticed. However, frequent alerts can lead to ``alert fatigue,'' where administrators become desensitised to notifications, potentially missing genuine threats amidst false positives \cite{mileski2022distributed}.

\textbf{Adaptive Billing Caps:} Adaptive billing caps allow application owners to set time-based spending limits, after which non-critical functions may be automatically disabled. This offers a fail-safe against runaway costs, although it risks disrupting services if legitimate usage unexpectedly spikes. Research highlights the need for intelligent capping mechanisms that can differentiate between malicious and legitimate usage patterns, ensuring that essential services remain operational even if billing limits are approached \cite{kelly2023dowts, ortega2024generation}.

\textbf{Native Cloud Provider Monitoring Tools:} Ben-Shimol et al. \cite{benshimol2025} present an unsupervised deep learning-based threat detection model aimed at identifying compromised functions in serverless cloud environments, including attacks such as Denial of Wallet (DoW). Unlike prior solutions requiring infrastructure modifications or third-party services, this model leverages only native cloud provider monitoring tools (e.g., AWS CloudTrail, CloudWatch, and X-Ray) to maintain full compatibility with managed serverless platforms. The detection framework captures post-exploitation behavioural anomalies by analysing high-granularity native logs and identifying deviations from learned execution baselines. A key innovation lies in its application-agnostic and threat-agnostic design, which supports broad scalability and ease of deployment across various serverless architectures. The model was evaluated in an AWS-based testbed across two open-source serverless applications, simulating DoW, data leakage, and permission misuse scenarios. It achieved successful detection of all attack types with a negligible false alarm rate of 0.003, highlighting its precision and robustness in high-noise, event-driven environments. The approach is particularly well-suited for organisations seeking to implement continuous anomaly monitoring without violating cloud service provider policies.

These mitigation strategies collectively offer a multi-faceted approach to defending against DoW attacks. However, their effectiveness varies depending on the specific type of DoW attack encountered. For instance, while rate limiting is ideal for Blast DDoW, it may have a limited impact on low-intensity attacks. Furthermore, the cost of implementing these strategies can be substantial, and organisations must weigh these costs against the potential financial impact of unmitigated DoW incidents. The need for more dynamic, context-sensitive mitigation systems is evident, prompting further research into adaptive methods that can respond to evolving attack tactics in real-time.

\begin{table*}[t]
\centering
\caption{Comparison of Denial of Wallet (DoW) Mitigation Strategies}
\label{tab:mitigation}
\renewcommand{\arraystretch}{1.2}
\resizebox{0.99\textwidth}{!}{%
\begin{tabular}{|p{3.6cm}|p{3.2cm}|p{3.2cm}|p{3.2cm}|p{4cm}|}
\hline
\textbf{Mitigation Strategy} & \textbf{Primary Function} & \textbf{Effective Against} & \textbf{Strengths} & \textbf{Limitations} \\
\hline
API Rate Limiting and Throttling & Restricts API calls per time unit; slows invocation during spikes & Blast DDoW & Effective burst control; immediate mitigation & May block legitimate high-traffic periods \\
\hline
Execution Time Limits & Terminates functions exceeding specified execution time & Slow-execution DoW tactics & Prevents cost inflation from long-running functions & Hard to calibrate for variable workloads \\
\hline
Cost Management Alerts & Notifies users when financial thresholds are exceeded & Continual Inconspicuous DDoW & Real-time cost visibility; early detection & Alert fatigue can reduce responsiveness \\
\hline
Adaptive Billing Caps & Caps daily/monthly spend and disables non-essential services & All types (as a failsafe) & Limits runaway costs; protects budgets & Risk of blocking legitimate usage spikes \\
\hline
Native Cloud Provider Monitoring Tools & Detects behavioural anomalies using native cloud monitoring & All types (via anomaly detection) & Scalable, non-intrusive, threat-agnostic detection & Relies on post-exploitation behaviour; not preventative \\
\hline
\end{tabular}
}
\end{table*}

\section{Machine Learning and AI Detection}

Machine learning (ML) and artificial intelligence (AI) have emerged as powerful tools in enhancing the detection and prevention of DoW attacks, especially in complex serverless environments where traditional monitoring methods fall short. The development of machine learning-based detection systems has advanced significantly, with approaches like DoWNet demonstrating the feasibility of using image classification techniques to identify malicious traffic patterns. Detecting DoW attacks, especially those that operate covertly over extended periods, has proven challenging. Early detection methods relied on identifying abnormal traffic volumes or invocation patterns, but these approaches were less effective against sophisticated DoW types like Continual Inconspicuous DDoW \cite{mileski2022distributed}.

\textbf{Gringotts:} The Gringotts system represents a significant breakthrough in DoW detection. By leveraging performance metrics at a granular level, Gringotts monitors parameters such as CPU usage, memory allocation, and function execution duration, analysing them in real-time to identify anomalies. The system employs statistical models, specifically Mahalanobis distance, to distinguish between normal and suspicious activity. This approach has proven highly effective, with Gringotts achieving a detection delay of just 1.86 seconds and an accuracy rate exceeding 95.75\%. Its low performance overhead—less than 1.1\%—also makes it suitable for deployment in high-demand environments where efficiency is paramount \cite{shen2022gringotts}.

Gringotts’s success underscores the value of advanced metrics in enhancing detection accuracy. Unlike conventional systems that rely on traffic volume alone, Gringotts considers a range of behavioural indicators, providing a comprehensive view of function performance. Comparisons with other detection mechanisms reveal that Gringotts’s real-time capabilities are particularly advantageous for applications with dynamic traffic patterns, where subtle deviations can signify malicious activity. While Gringotts is not a standalone solution, it exemplifies the potential of combining statistical analysis with real-time monitoring to address the unique challenges of DoW detection.

\textbf{DoWNet and CNNs:} DoWNet represents a novel application of deep learning in DoW detection. By converting request data into visual heatmaps, DoWNet uses convolutional neural networks (CNNs) to classify patterns as normal or suspicious. The heatmap representation captures temporal and spatial variations in traffic, allowing the model to detect subtle anomalies that may signal an ongoing DoW attack. Studies on DoWNet report an accuracy rate of 97.98\%, indicating the model’s effectiveness in differentiating between legitimate and malicious traffic patterns \cite{kelly2024downet}.

\textbf{FODWNN-DoWAD:} FODWNN-DoWAD \cite{renukadevi2025} is a novel detection framework for Denial of Wallet (DoW) attacks in serverless computing environments. The proposed framework integrates three core components: Pair Barracuda Swarm Optimization (PBSO) for feature selection, Deep Wavelet Neural Networks (DWNN) for capturing temporal and frequency-domain patterns, and the Hierarchical Learning-based Chaotic Crayfish Optimizer (HLCCO) for hyperparameter tuning. Experimental evaluations on a benchmark dataset yielded a detection accuracy of 99.05\%, outperforming existing methods in classification performance and computational efficiency. The use of wavelet-based neural architecture enhances detection in high-noise environments typical of transient serverless executions.

\textbf{FaaSMT:} FaaSMT is a lightweight intrusion detection framework designed to address the growing threat of Denial of Wallet (DoW) and business logic attacks in serverless computing environments \cite{li2025faasmt}. FaaSMT integrates Merkle Tree-based trust verification with adaptive task inlining to detect and prevent unauthorised function execution and communication anomalies. It tracks invocation chains using REST APIs and constructs Merkle Trees from SHA-256 hashes of function execution logs to ensure data integrity and provenance. Experimental evaluations on AWS using representative FaaS applications demonstrate that FaaSMT significantly improves security verification coverage while maintaining system responsiveness and low resource consumption.

\textbf{Other ML Models:} Beyond CNNs, researchers have explored other machine learning techniques for DoW detection, such as Random Forests and Support Vector Machines (SVMs). While these models are different to CNNs, they have proven effective in identifying specific types of DoW attacks when provided with structured, labelled datasets. Random Forests, for example, are adept at handling large datasets with multiple variables, making them useful in environments with diverse function types and varying traffic loads. However, studies suggest that these models may require significant feature engineering to achieve accuracy levels comparable to deep learning models like DoWNet \cite{shen2022gringotts}.

\textbf{Traditional Detection Models:} Traditional rule-based detection models often struggle to identify low-visibility DoW attacks like Continual Inconspicuous DDoW, which operate within normal traffic parameters. By contrast, ML-based approaches excel at recognising nuanced deviations in traffic patterns, making them highly suitable for detecting low-intensity, prolonged attacks.

\begin{table*}[t]
\centering
\caption{ML and AI Solutions for DoW Attack Mitigation}
\label{tab:ml_detection}
\renewcommand{\arraystretch}{1.2}
\resizebox{0.99\textwidth}{!}{%
\begin{tabular}{|p{2.8cm}|p{1.3cm}|p{0.8cm}|p{2.8cm}|p{3cm}|p{3.2cm}|p{3.2cm}|}
\hline
\textbf{Solution} & \textbf{Citation} & \textbf{Year} & \textbf{Technique} & \textbf{Detection Scope} & \textbf{Accuracy / Performance} & \textbf{Strengths / Weaknesses} \\
\hline
Gringotts & \cite{shen2022gringotts} & 2022 & Mahalanobis Distance & Behavioural anomalies from performance metrics & $>$95.75\% accuracy, 1.86s delay, $<$1.1\% overhead & Fast, accurate, low overhead; not standalone \\
\hline
DoWNet (CNN) & \cite{kelly2024downet} & 2024 & Convolutional Neural Networks & Temporal-spatial heatmap classification & 97.98\% accuracy & Captures subtle patterns; heatmap conversion dependency \\
\hline
FODWNN-DoWAD & \cite{renukadevi2025} & 2025 & DWNN + PBSO + HLCCO & Frequency-domain + temporal detection & 99.05\% accuracy & Highly optimised; unknown real-time feasibility \\
\hline
FaaSMT & \cite{li2025faasmt} & 2025 & Merkle Tree + Task Inlining & Log-based execution chain integrity & Qualitative results only & Lightweight; no explicit accuracy/benchmark data \\
\hline
Random Forests / SVMs & - & - & Classical ML models & Structured function-based traffic patterns & Effective with engineered features & Needs feature engineering; weaker than CNNs \\
\hline
Traditional Detection & - & - & Rule-based thresholds & High-volume traffic spikes & Misses stealthy DoW & Simple; ineffective on low-intensity prolonged attacks \\
\hline
\end{tabular}
}
\end{table*}

\section{Discussion}

To effectively mitigate the multifaceted threat posed by Denial of Wallet (DoW) attacks in cloud-native and serverless environments, it is essential to match each specific attack type with the most appropriate machine learning (ML) and artificial intelligence (AI)-based detection strategies. Table~\ref{tab:ai_mapping} provides a strategic mapping between three core DoW attack variants—Blast DDoW, Continual Inconspicuous DDoW, and Background Chained DDoW—and the ML/AI solutions most suited to their detection and containment. Each attack type exploits different architectural and operational characteristics of serverless platforms, thus necessitating differentiated, context-aware mitigation approaches.

\begin{table}[htbp]
\centering
\caption{Recommended ML/AI Solutions per DoW Attack Type}
\label{tab:ai_mapping}
\renewcommand{\arraystretch}{1.2}
\resizebox{0.999\linewidth}{!}{%
\begin{tabular}{|p{3.5cm}|p{4.8cm}|}
\hline
\textbf{Attack Type} & \textbf{Recommended AI/ML Solutions} \\
\hline
Blast DDoW & Gringotts, FODWNN-DoWAD, Traditional Detection Models \\
\hline
Continual Inconspicuous DDoW & DoWNet (CNN), Random Forests, SVMs \\
\hline
Background Chained DDoW & FaaSMT, FODWNN-DoWAD \\
\hline
\end{tabular}
}
\end{table}

\textbf{Blast DDoW attacks}—characterised by their high-intensity and short-duration invocation patterns—aim to generate immediate and severe financial strain through rapid exploitation of serverless auto-scaling. To detect and respond to these cost-amplifying bursts, models such as Gringotts are well suited due to their real-time monitoring of CPU, memory, and execution metrics combined with statistical anomaly detection (e.g., Mahalanobis distance). In addition, FODWNN-DoWAD, with its deep wavelet neural network architecture and optimised feature selection, provides high detection accuracy (99.05\%) and robust performance under bursty, noisy input conditions. While less sophisticated, traditional detection models still play a complementary role, as their threshold-based alerts can be useful for catching volume-based anomalies.

\textbf{Continual Inconspicuous DDoW attacks} represent a more stealthy and prolonged threat, with attackers issuing a low but steady stream of invocations over an extended period. These attacks blend malicious traffic into normal background workloads, making them challenging to detect using volume-based metrics alone. For such scenarios, DoWNet, which leverages convolutional neural networks (CNNs) to analyse temporal-spatial heatmaps of function invocation patterns, is highly effective. Its ability to detect subtle deviations within otherwise legitimate-looking traffic allows it to uncover slow-burn attacks that evade traditional tools. Similarly, Random Forests and SVMs can be valuable in these contexts when paired with structured, labelled datasets and appropriate feature engineering, enabling detection of nuanced anomalies over time.

\textbf{Background Chained DDoW attacks} exploit the event-driven and interconnected nature of modern serverless architectures by creating cascading invocation chains across multiple functions. Here, FaaSMT stands out due to its Merkle Tree-based trust verification and task inlining mechanisms, which allow it to trace call chains and identify unauthorised or anomalous execution paths. Additionally, FODWNN-DoWAD remains effective in this context, as it can detect abnormal behaviour patterns even within complex microservice orchestrations.

Overall, this mapping illustrates the necessity for precision-aligned detection strategies that reflect the operational subtleties of each DoW attack variant.

\section{Research Gaps and Future Direction}

While significant advancements have been made in detecting and mitigating DoW attacks, the field faces several ongoing challenges. These challenges are compounded by the dynamic nature of serverless environments and the constantly evolving tactics of attackers.

\textbf{Limited Access to Real-World Data:} The lack of real-world DoW attack datasets remains a major obstacle for researchers. Organisations are often reluctant to disclose incidents of financial exploitation, leading to a scarcity of empirical data that could inform detection model training and validation. The use of synthetic data generation tools like DoWTS partially mitigates this issue, but synthetic datasets cannot fully capture the variability and noise present in real-world traffic. Developing collaborative frameworks for data sharing, anonymised to protect organisational privacy, could advance research by providing more representative data for analysis \cite{kelly2023dowts, ortega2024generation}.

\textbf{Adapting to Evolving Attack Tactics:} As detection mechanisms improve, attackers may adapt by developing more sophisticated DoW techniques that evade traditional defences. For instance, future DoW attacks could leverage distributed botnets to orchestrate coordinated low-intensity attacks across multiple endpoints, making them harder to detect with single-point monitoring systems. Research is beginning to explore adaptive detection models capable of learning from new attack patterns, but implementing these systems at scale remains challenging due to computational and financial constraints \cite{kelly2024downet}.

\textbf{Policy and Billing Model Reforms:} There is growing recognition of the need for policy changes among cloud providers to enhance security against DoW attacks. Suggested reforms include adaptive billing models that can dynamically adjust to usage anomalies, flagging potential DoW activity before costs escalate. Some researchers propose that cloud providers offer built-in detection and mitigation services as part of their platform, reducing the burden on individual application owners. However, implementing such changes poses challenges for providers, as it may require significant modifications to existing billing structures and service agreements \cite{shen2022gringotts}.

\textbf{Collaborative Research and Open-Source Tools:} The success of tools like DoWTS has demonstrated the value of open-source platforms for advancing DoW research. By making these tools freely accessible, researchers can build on shared resources to develop more robust detection and mitigation systems. Moving forward, fostering an open-source ecosystem around DoW research could accelerate innovation and enable researchers to address emerging challenges more effectively. Collaboration between academia, industry, and cloud providers is essential for developing solutions that are both practical and scalable \cite{kelly2023dowts, ortega2024generation}.

As DoW threats continue to evolve, future research will need to address the limitations of current detection and mitigation models. Key areas of focus include refining machine learning algorithms to minimise false positives, exploring decentralised detection architectures for distributed DoW attacks, and developing policies that incentivise cloud providers to implement proactive billing safeguards.

\section{Conclusion}

The Denial of Wallet (DoW) attack has become as a significant threat in the context of serverless computing, where the benefits of scalability and pay-as-you-go billing models are exploited to cause financial harm. As the literature reveals, DoW attacks differ fundamentally from traditional Denial of Service (DoS) attacks by shifting the objective from resource exhaustion to financial depletion. The evolution of research from early awareness to advanced detection and mitigation illustrates the urgency with which the cybersecurity community has addressed this growing threat. Through developments in classification, simulation tools like DoWTS, and advanced detection mechanisms such as Gringotts and DoWNet, researchers have created a robust foundation for identifying and managing DoW attacks.

The thematic analysis presented highlights the breadth of DoW tactics, from rapid, high-frequency attacks to subtle, prolonged exploitation methods. Each approach demands specialised detection and mitigation strategies, underscoring the complexity of financially motivated cyber threats in serverless environments. Despite substantial progress, gaps remain, particularly in the availability of real-world datasets and the adaptability of current systems to new attack forms. Mitigating these challenges will require a collaborative approach across academia, industry, and cloud providers to establish standardised policies, develop adaptive billing systems, and expand open-source resources.

This review underscores the necessity for continued innovation in detection and defence methodologies. Machine learning models, particularly deep learning approaches like those in DoWNet, demonstrate promise in refining detection accuracy, but challenges persist in implementing cost-effective, scalable defences for all serverless users.

As serverless architectures continue to change, future research must adapt to the evolving tactics of DoW attackers, focusing on sustainable, responsive solutions that protect application owners from financial exploitation.

\end{document}